\documentclass{aa}
\vspace{-2truecm}
\usepackage{epsfig}
\begin{document}
\def\kms{km~s$^{-1}${}}

\title{The detection of stellar velocity dispersion drops in the central 
regions of five isolated Seyfert spirals. 
\thanks{Based on observations made with the WHT operated on the island of
La Palma by ING in the Spanish Observatorio del Roque de Los Muchachos
of the Instituto de Astrof\'\i sica de Canarias}
}
\author{
  I.~M\'arquez \inst{1}
\and
  J.~Masegosa \inst{1}
\and
  F.~Durret \inst{2}
\and
  R.M.~Gonz\'alez Delgado  \inst{1}
\and
  M.~Moles \inst{1}
\and
  J.~Maza \inst{3}
\and
  E.~P\'erez \inst{1}
\and
  M.~Roth \inst{4}}
\offprints{I. M\'arquez (\sl{isabel@iaa.es}) }
\institute{
	Instituto de Astrof\'\i sica de Andaluc\'\i a (C.S.I.C.), 
Apartado 3004 , E-18080 Granada, Spain
\and
	Institut d'Astrophysique de Paris, CNRS, 98bis Bd Arago, 
F-75014 Paris, France 
\and 
	Departamento de Astronom\'\i a, Universidad de Chile, Casilla 36D, 
Santiago, Chile
\and
    Observatories of the Carnegie Institution of Washington, 813 Barbara 
Street, Pasadena, CA91101
}

\date{Received / Accepted } \authorrunning{M\'arquez et al.}

\titlerunning{Stellar velocity dispersion drops in isolated Seyfert spirals}

\abstract { We analyze the kinematics of the central regions of five
isolated Seyfert spiral galaxies from the DEGAS sample 
(four with new data presented in this paper, IC~184, UGC~3223,
NGC~2639, NGC~6814, and NGC~6951 from our previous data), by
using long slit spectroscopy in the CaII triplet range (at
$\approx$ 8600\AA) obtained with a 4m-class telescope.  A drop of the
velocity dispersions in the innermost $\pm$(1-3)~arcsec is observed in
four of them, and hinted in the
remaining galaxy (NGC~6814). The available HST images for our sample together
with another nine galaxies with reported velocity dispersion drops,
are also used to investigate the presence of morphological inner
structures at the scales of the kinematical drops.  Evidence
for disk-like shapes is found in 12 out of the 14 cases. The only
exceptions are NGC~6814 and NGC~6951.  Existing N-body simulations
including stars, gas and star formation predict that such a drop is
most probably due to a young stellar population born from dynamically
cold gas accreted in a circumnuclear disk formed during an episode of
central gas accretion driven by a bar.  The equivalent widths of the
Calcium triplet lines for our 5 galaxies have been measured. Even
if the profiles could be formally consistent with constant EW(CaT)
values, they seem to indicate the presence of a local maximum
in the regions corresponding spatially to the drops; 
if confirmed, this would imply the presence of a
different stellar population, whose properties could help constraining
the models.  
\keywords {galaxies: spiral -- galaxies: kinematics and
dynamics -- galaxies: structure -- galaxies: interaction -- galaxies:
individual: IC~184, UGC~3223, NGC~2639, NGC~6814, NGC~6951} }
\maketitle

\section{Introduction}

The existence of non-axisymmetric components of the galactic potential
has been frequently invoked as an efficient way to transport gas from
the galaxy scale down to the nucleus to fuel the active galactic
nucleus (AGN). In particular, the shocks and gravitational torques
induced by a galactic bar provide the non axisymmetric potential
invoked in theoretical works (Simkin et al. 1980, Shlosman et
al. 1989, Barnes \& Hernquist 1991) to make the gas lose angular
momentum and therefore facilitate the fueling mechanism. Whether there
is no preference for Seyfert nuclei to occur in barred galaxies
(Heckman 1980, Simkim et al. 1980, Moles et al. 1995, McLeod \& Rieke
1995, Mulchaey \& Regan 1997) or whether an excess of bars among
Seyferts exists, 
is still an open
issue. Nevertheless, in addition to the large scale bar, another
mechanism is needed to drive the gas to the very central regions; one
such mechanism is that of nested bars (Shlosman et al. 1989;
Friedli \& Martinet 1993; Combes 1994; Heller \& Shlosman 1994), which
has been suggested to fuel molecular gas into an intense central
starburst in NGC 2782 (Jogee et al. 1999) and in a Seyfert 2 galaxy,
Circinus (Maiolino et al. 2000). However, the analysis of the HST images of
Seyferts (Regan \& Mulchaey 1999, Martini \& Pogge 1999, P\'erez et
al. 2000, Martini et al. 2003a, 2003b) has shown that
such nuclear bars are not ubiquitous, and nuclear spirals are proposed
to be one of the channels to feed gas into the central engine.

Stellar kinematical profiles of four double barred galaxy hosts of
AGNs obtained with the VLT and ISAAC have recently
shown the existence of decoupled kinematics in the nuclear region, as
well as a drop of the velocity dispersion in three of them (Emsellem
et al. 2001). Although their models could not properly reproduce this
drop, Emsellem et al. suggested that it was due to a transient cold
nuclear disk fueled by gas inflow along the bar, which has recently
formed new stars. Models by Wozniak \& Michel-Dansac (2003) have
recently succeeded in accounting for this drop in velocity dispersion,
which indeed seems to be due to a young stellar population born
from dynamically cold gas accreted in a circumnuclear disk. The
presence of such disks can be directly related to the circumnuclear
structures required to drive the gas to the very central regions, as
nuclear (nested) bars or nuclear spirals (see references above).

In most of these studies the environmental effects are in general not
considered when studying the properties of Seyfert galaxies and/or
comparing them with those of non-active ones (but see Moles et
al. 1995). However, the properties of spiral galaxies can be modified
even in mild interactions, as shown by M\'arquez et al. (2002) for a
sample of 111 galaxies ranging from truly isolated to mildly
interacting spirals. Therefore, we decided to study the properties of
isolated Seyfert spirals, and to compare them with a control sample of
isolated non-active spirals; the main result is that the active and
non-active samples are equivalent in the studied
properties: large scale disks, bulges and bars have the same
properties, as derived by the analysis of their NIR images (M\'arquez
et al. 1999, 2000). To further delimit eventual
differences/similarities and to put some constraints on the possible
mechanisms taking place in the circumnuclear regions, kinematical
information is needed; therefore, long slit spectra along several
position angles were obtained for a number of our sample galaxies, to
get both gas (2/3 of the whole sample) and stellar kinematics (for a
subsample). The results of the overall kinematical analysis together
with the information on metallicities provided by the analysis of the
HII regions will be published separately (M\'arquez et al. 2003). High
spectral and spatial resolution spectra in the CaT region have been
obtained for a total of eight Seyfert galaxies.  In four of them
we have found drops in the stellar velocity dispersion of the central
regions, together with a hint of such a drop in NGC~6814,
comparable to those reported by Emsellem et al. (2001).

Sect.~2 is devoted to the presentation of the observations, data
reduction and analysis. The results of the analysis are reported in
Sect.~3, and the discussion and conclusions in Sect.~4.

\section{Observations, Data Reduction and Analysis}

The DEGAS sample includes 33 isolated galaxies, 17 with an
AGN and 16 without an AGN taken as a control sample.
The four galaxies reported here all have an AGN and their
infrared properties have already been described (M\'arquez
et al. 1999). As a complement to these infrared images, we have
retrieved the HST images for these four galaxies and present them
here. All were taken with the F606W filter.

The spectroscopic data for the gas kinematics were collected with the
ISIS Spectrograph at the 4m William Herschel Telescope (WHT) in La
Palma (Spain). The setup and main characteristics of the observations
are given in Table~\ref{obs}. The spectral and spatial samplings were
0.39\AA/pixel and 0.36 arcsec/pixel, respectively, and the slit
width 1.03 arcsec. The spectral resolution given in Table~\ref{obs} 
has been measured as the FWHM of a gaussian fit to several unblended sky 
lines.

\begin{table}[]
\caption[]{Detailed log of the spectroscopic observations}
\label{obs}
\begin{flushleft}
\begin{scriptsize}
\begin{tabular}{lc c c r r r}
\hline
\hline
\noalign{\smallskip}
\multicolumn {1}{l}{Galaxy}
& \multicolumn{1}{c}{Date}
& \multicolumn{1}{r}{Spectral}
& \multicolumn{1}{r}{$<$Seeing$>$}
& \multicolumn{1}{r}{PA}
& \multicolumn{1}{c}{Exp.}
& \multicolumn{1}{c}{Spect.}\\
\multicolumn {1}{l}{}
& \multicolumn{1}{c}{}
& \multicolumn{1}{r}{range (\AA)}
& \multicolumn{1}{r}{(arcsec)~}
& \multicolumn{1}{r}{(\degr)}
& \multicolumn{1}{c}{(sec)}
& \multicolumn{1}{c}{resol.}\\
\noalign{\smallskip}
\hline\noalign{\smallskip}
IC 184     & 1999  & 8476-8872 & 1.5 &   7 & 7200 & 0.71 \\     
           & 1999  & 8476-8872 & 1.5 &  97 & 5400 & 0.71 \\
UGC 3223   & 1999  & 8476-8872 & 1.5 &  80 & 5400 & 0.71 \\          
NGC 2639   & 1999  & 8476-8872 & 1.5 &  45 & 5400 & 0.68 \\          
           & 1999  & 8476-8872 & 1.5 & 135 & 5400 & 0.71 \\          
NGC 6814   & 1996  & 8505-8881 & 1.2 &  30 & 3600 & 0.71 \\
           & 1996  & 8505-8881 & 1.2 & 120 & 5400 & 0.78 \\
\hline
\end{tabular}
\noindent
\end{scriptsize}
\end{flushleft}
\end{table}

\begin{table}[]
\caption[]{Details on HST retrieved images}
\label{tabhst}
\begin{flushleft}
\begin{scriptsize}
\begin{tabular}{l l l r c c}
\hline
\hline
\noalign{\smallskip}
\multicolumn {1}{l}{Galaxy}
& \multicolumn{1}{l}{AGN$^*$}
& \multicolumn{1}{l}{Band}
& \multicolumn{1}{c}{Exp.}
& \multicolumn{1}{c}{Prog.}
& \multicolumn{1}{c}{PI}\\
\multicolumn {1}{l}{}
& \multicolumn{1}{c}{}
& \multicolumn{1}{c}{}
& \multicolumn{1}{c}{(sec)}
& \multicolumn{1}{c}{Nr.}
& \multicolumn{1}{c}{}\\
\noalign{\smallskip}
\hline\noalign{\smallskip}
IC 184     & Sy2 & F606W & 500 & 5479 & M. Malkan\\ 
UGC 3223   & Sy1 & F606W & 500 & 5479 & M. Malkan\\     
NGC 2639   & Sy1.9 & F606W & 500 & 5479 & M. Malkan\\     
NGC 6814   & Sy1.5 & F606W & 500 & 5479 & M. Malkan\\     
\hline
NGC 1395   & Sy1.8 & F606W & 1680& 8597 & M. Regan \\
NGC 1808   & Sy2 & F814W &  160& 6872 & J. Flood \\
NGC 3412   & -- & F606W&   160& 5446 & G. Illingworth\\
NGC 3623   & L=LINER & F606W&   160& 5446 & G. Illingworth\\
NGC 3627   & L/Sy2 & F606W&   560& 8597 & M. Regan\\
NGC 4303   & Sy2 & F814W&   460& 9042 & S. Smartt\\
NGC 4477   & Sy2 & F606W&   160& 5446 & G. Illingworth\\
NGC 4579   & L/Sy1.9& F658N&  1400& 6436 & D. Maoz\\
NGC 4725$^{**}$   & Sy2 & F606W&   560& 8597 & M. Regan\\
NGC 6503   & L & F814W&  2600& 8602 & A. Filippenko\\
\hline
\end{tabular}
\null
\noindent
$^{*}$ Taken from NED\\ 
$^{**}$ Diffraction spikes impede a reliable determination of the
hinted central elongated structure\end{scriptsize}
\end{flushleft}
\end{table}

\begin{figure*}[]
\begin{picture}(300,240)
\end{picture}
\vspace{3.5cm}
\caption{IC~184. Top left: velocity curve of the gas (open circles,
from M\'arquez et al. 2003) and stars (black circles) along PA=7\degr.
Top middle: FWHM of the stellar component along PA=7\degr.  Top right:
FWHM of the stellar component along PA=97\degr. Bottom left: HST image
of IC~184 in the F606W band with the two slits superimposed; the
ellipses correspond (PA and ellipticities) to the two bars detected in
the infrared (M\'arquez et al. 1999). Bottom middle: sharp divided
image of the center.}
\label{fig:i184}
\end{figure*}

Standard IRAF\footnote{IRAF is the Image Analysis and Reduction
Facility made available to the astronomical community by the National
Optical Astronomy Observatories, which are operated by the Association
of Universities for Research in Astronomy (AURA), Inc., under contract
with the U.S. National Science Foundation.} procedures were used for
the reduction of the spectroscopic data, following the standard steps
of bias subtraction, flat field correction, wavelength calibration with
a CuNe lamp observed before and after the target, atmospheric
extinction correction, and flux calibration using spectroscopic
standards observed through an 8 arcsec wide slit. The sky background
level was determined taking median averages over two strips on either
side of the galaxy signal.  We used cross-correlation techniques as
described in P\'erez et al. (2000) and references therein, to extract
the kinematic information, using as templates for the
cross-correlation the observed spectra of several stars.  The errors
refer to the dispersion in the velocity shifts measured with respect 
to the different template stars.
For clarity, error bars in the stellar velocity
distributions are not plotted, but they typically amount to $\pm$50 \kms,
with median values of $\pm$35 \kms~ for NGC~6951, and up to $\pm$70 \kms~ for
NGC~2639. The systemic velocity has been chosen as that of the
section with the maximum continuum level, except for IC~184, for which
the kinematical center (the point that allows the best symmetrization
of the rotation curve) is shifted by about 0.8 arcsec to the South with 
respect to the continuum maximum.

The Calcium triplet equivalent width has been measured following the
method described by Terlevich et al. (1990, hereafter, TDT90), which
uses two specific continuum bands on either side of the CaT region,
after Doppler shift corrections for each section.  Due to a
shorter spectral range coverage than TDT90, slight modifications of
the central wavelength of the ``blue'' continuum were adopted for all
the galaxies except NGC~2639. For the ``red'' continuum, we had to use
much shorter wavelengths than TDT90.  The resulting equivalent widths
for the inner regions of each galaxy (see Sect.~3) are given in Table
\ref{ew}.  In the last two columns the central rest frame wavelength
of each continuum band is given (the width used is 15 \AA).  The
differences in the continuum bands used may be the reason why we
measure systematically lower equivalent widths (EW) than those
reported in TDT90 for the same types of active galaxies.  As TDT90
pointed out, to obtain conclusions on the stellar population based on
CaII triplet measurements, exactly the same method has to be used. For
our data the results are specially tricky due to the fact that the TiO
and VO bands contaminate the ``red'' continuum. In order to quantify
the uncertainty due to the choice of the continuum level, we have used
three different ``red'' continua and measured the corresponding EW;
the error bars in Fig. \ref{ewCaT} are calculated as the dispersion of
these three values, and therefore represent lower limits to the
true error bars. The minimum uncertainty obtained for the best cases
with a flat continuum is 0.5 \AA\ but in most of the galaxies the
uncertainty can be larger than 1.5 \AA.

\begin{figure*}[]
\caption{UGC~3223. Top left: velocity curve of the gas (open circles, 
from M\'arquez et al. 2003)
and stars (black circles) along PA=80\degr.
Top right: FWHM of the stellar component along PA=80\degr.
Bottom left:
HST image of UGC~3223 in the F606W band with the two slits superimposed;
the ellipses correspond (PA and ellipticities) to the two bars
detected in the infrared (M\'arquez et al. 1999). Bottom right: sharp
divided image of the center.}
\label{fig:u3223}
\end{figure*}

CaT equivalent widths have also been measured using the method by
Cenarro et al. (2001). The comparison between this method and that of TDT90
has shown that, in the case of dilution due to an important underlying
old stellar population, TDT90 seem to be less sensitive to the contamination
by TiO and VO lines. Therefore, since our sample galaxies are all
early type spirals, we have adopted the TDT90 method.

\begin{table}[]
\caption[]{CaT Equivalent widths}
\label{ew}
\begin{flushleft}
\begin{scriptsize}
\begin{tabular}{l r c c c c c}
\hline
\hline
\noalign{\smallskip}
\multicolumn {1}{l}{Galaxy}
& \multicolumn{1}{c}{PA}
& \multicolumn{1}{c}{EW}
& \multicolumn{1}{c}{EW}
& \multicolumn{1}{c}{EW}
& \multicolumn{1}{c}{``blue''}
& \multicolumn{1}{c}{``red''}\\
\multicolumn {1}{l}{}
& \multicolumn{1}{c}{}
& \multicolumn{1}{c}{Ca2}
& \multicolumn{1}{c}{Ca3}
& \multicolumn{1}{c}{CaII}
& \multicolumn{1}{c}{cont.}
& \multicolumn{1}{c}{cont.}\\
\noalign{\smallskip}
\hline\noalign{\smallskip}
IC 184     & 7 & 3.5 (0.3) & 2.8 (0.6) & 6.3 (1.8) & 8458& 8680 \\ 
           & 97& 4.0 (0.3) & 2.7 (0.4) & 6.7 (0.6) & 8458& 8680 \\    
UGC 3223   & 80& 3.5 (0.1) & 2.8 (0.3) & 6.3 (0.7) & 8458& 8700 \\     
NGC 2639   & 45& 3.0 (0.1) & 2.6 (0.1) & 5.7 (0.3) & 8455& 8700 \\     
           &135& 3.5 (0.1) & 2.6 (0.2) & 6.1 (0.5) & 8455& 8700 \\     
NGC 6814   & 30& 3.5 (0.2) & 2.5 (0.5) & 6.1 (1.3) & 8484& 8700 \\     
           &120& 3.5 (0.2) & 2.5 (0.5) & 6.0 (1.3) & 8484& 8700 \\     
NGC 6951   & 48& 3.7 (0.1) & 3.3 (0.1) & 7.0 (0.3) & 8482& 8750 \\     
           &138& 3.4 (0.1) & 3.1 (0.3) & 6.6 (0.7) & 8482& 8750 \\     

\hline
\end{tabular}
\noindent
\null
Ca2 = CaT line at 8542\AA\\
Ca3 = CaT line at 8662\AA\\
EW(CaII) = EW(Ca2) + EW(Ca3)
\end{scriptsize}
\end{flushleft}
\end{table}

For our five sample galaxies and for nine of the eleven galaxies with
velocity dispersion drops reported in the literature (see Sect.~4) the
available archival WFPC2 HST\footnote{Based on observations collected with the
NASA/ESA HST, obtained at STScI which is operated by AURA Inc., under
NASA contract NAS5-26555} optical images
(see Table \ref{tabhst} for details) have been used to identify
features which, according to numerical models, could
possibly be related to the kinematics.  To trace the presence of
morphological features in the innermost few arcseconds, both raw and
sharp--divided images\footnote{The result of dividing the original
images by median--filtered ones.} have been used.  Sharp--divided
images are very useful to trace asymmetries in the light distribution,
such as bars, spiral arms, dust lanes and rings. This technique allows
the subtraction of the diffuse background in a very convenient way to
look for subtle, small-scale variations and discuss the possible
presence of both dust extinguished and more luminous regions (Sofue et
al. 1994; M\'arquez \& Moles 1996; Erwin \& Sparke 1999; Laine et
al. 1999; M\'arquez et al. 1999; Erwin \& Sparke 2002).

\begin{figure*}[]
\begin{picture}(300,240)
\end{picture}
\vspace{3.5cm}
\caption{NGC~2639. Top left: velocity curve of the gas (open circles,
from M\'arquez et al. 2003) and stars (black circles) along
PA=45\degr.  Top middle: FWHM of the stellar component along
PA=45\degr.  Top right: FWHM of the stellar component along
PA=135\degr. Bottom left: HST image of NGC~2639 in the F606W band with
the two slits superimposed; the ellipses correspond (PA and
ellipticities) to the two bars detected in the infrared (M\'arquez et
al. 1999). Bottom middle: sharp divided image of the center.}
\label{fig:n2639}
\end{figure*}

\section{Results}

The kinematical data and the images used in the analysis of the data are
displayed in Figs.~\ref{fig:i184} to \ref{fig:n6814}. The points
corresponding to gas kinematics are given for comparison (for further
details on the gas kinematics see M\'arquez et al. 2003). The slit
positions (measured from North to East) are drawn on the figures,
together with the bar sizes and ellipticities as determined in
M\'arquez et al. (1999), for the primary bar in single-barred galaxies
(NGC~6814 and NGC~6951) and for primary and secondary (inner) bars in
double-barred ones (IC~184, NGC~2639 and UGC~3223). The results for
the CaII triplet EW measurements of each galaxy are presented in Table
\ref{ew}. Since we are interested in searching for the properties of
stellar populations in the regions where the dips have been detected,
the values given in Table \ref{ew} refer to the whole region occupied
by the drops.
 
\subsection{IC~184}

The rotation curves along PA=7\degr\ (major axis)
(Fig.~\ref{fig:i184}) and 97\degr~ show that the rotation of the gas
and stars are strongly decoupled along PA=7\degr~ within $\pm$4 arcsec
of the center, with a central region of stellar counter-rotation
within 1 arcsec, which is also visible along PA=97\degr~(not shown
here).  In order to know if such kinematics are due to the
presence of a counter rotating core, as it is the case for NGC~4621
(Wernli et al. 2002) integral field spectroscopy is required. A small
dip of about 20 \kms~ with respect to the maximum value is observed in
the velocity dispersion in the innermost $\pm$2 arcsec. The stellar
velocity distribution along PA=97\degr~ shows structure within the
innermost $\pm$9 arcsec, with a wavy central shape similar to that
observed along PA=7\degr. This extension corresponds to that of the
primary bar, which extends along PA=170\degr~ (M\'arquez et al. 2000),
i.e. only 17\degr~ from the disk major axis. We note that the
secondary bar extends $\pm$4 arcsec along PA=30\degr~ (M\'arquez et
al. 2000) and therefore the {\large} velocity difference between the
stellar and gas components in this region is most likely
due to the non-circular motions caused by the
secondary bar. EW(CaT) shows a slight tendency for central values to
be higher along both PAs, particularly along PA=97\degr (Fig. \ref{ewCaT}a).

The HST image of IC~184 (Fig.~\ref{fig:i184}) shows a structure in the
innermost 1~arcsec elongated along PA $\approx$ 50\degr, very close to
the position angle of the inner bar in the same image (PA
$\approx$ 40\degr); although this feature is detected in a much
smaller region, its PA is consistent with that of the small inner bar
already detected in the infrared (M\'arquez et al. 1999). A structure
resembling two straight dust lanes may be seen in the sharp-divided
image running along about PA $\approx$ 45\degr, below the center to the SW 
and above the center to the NE, respectively.

\begin{figure*}[]
\begin{picture}(300,240)
\end{picture}
\vspace{3.5cm}
\caption{NGC~6814. Top left: velocity curve of the gas (open circles,
from M\'arquez et al. 2003) and stars (black circles) along
PA=120\degr.  Top middle: FWHM of the stellar component along
PA=120\degr.  Top right: FWHM of the stellar component along
PA=30\degr. Bottom left: HST image of NGC~6814 in the F606W band with
the two slits superimposed; the ellipse corresponds (PA and
ellipticity) to the bar detected in the infrared (M\'arquez et
al. 1999). Bottom middle: sharp divided image of the center.}
\label{fig:n6814}
\end{figure*}

\subsection{UGC~3223}

The rotation curve of UGC~3223 along PA=80\degr\ (very close to the bar
major axis, PA=75\degr), although rather noisy, shows that the
overall rotation of the gas follows, within the error bars, that
of the stars (Fig.~\ref{fig:u3223}). Positive stellar velocities
are systematically lower (down to 80 \kms) than their negative
counterparts for radii between 2 and 20 arcseconds. 
A small dip ($\approx$ 10 \kms) may be observed in the
velocity dispersion in the innermost $\pm$ 1 arcsec, but needs to
be confirmed with data of higher signal to noise. In spite of the low
S/N ratio of the spectra, an 
increase in EW(CaT)
is detected in the innermost $\pm$1 arcsec, corresponding to the
extension of the detected dip (see Fig. \ref{ewCaT}b).

The HST image of UGC~3223 (Fig.~\ref{fig:u3223}) shows an 
asymmetric structure in the innermost 1~arcsec elongated mainly along
PA=80\degr, and more extended to the West, consistent with that
of the small inner bar already detected in the infrared (M\'arquez et
al. 1999).  Another extension can be
seen to the NE along PA$\sim$55\degr, resembling what could be an incomplete
ring at r=0.9 arcsec.

\subsection{NGC 2639}

NGC~2639 is a galaxy optically classified as unbarred, but barred at
NIR wavelengths (M\'arquez et al. 1999). The stellar and gas rotation
curves of NGC~2639 along PA=135\degr~(disk and main bar major axis,
not shown here) show that both components have different velocities
(up to 100 \kms) in the region inside the bar (inner
$\pm$8 arcsec) (M\'arquez et al. 2003).  The gaseous velocity
distribution along PA=0\degr~ shows departures of 50 \kms from pure rotation
in the region within r$<$4 arcsec (M\'arquez et
al. 2003).  Along PA=45\degr~ the velocity distribution is quite flat,
as expected along a position angle close to the minor axis, whereas
clear non-circular motions are traced by the gas kinematics in the
central $2-3$ arcsec radius (Fig.~\ref{fig:n2639}). A velocity
dispersion drop of $\approx$ 20 \kms\ is detected along the two PAs in
the innermost $\pm$2~arcsec, being more clearly detected along
PA=45\degr~(Fig.~\ref{fig:n2639}).
A hint of an increase in EW(CaT) is detected along
PA=45\degr and 135\degr (Fig. \ref{ewCaT}c).

The HST image of NGC~2639 (Fig.~\ref{fig:n2639}) shows a structure
elongated along PA=135\degr~ in the innermost $\pm$1~arcsec
(consistent with the thin inner bar detected in the infrared,
M\'arquez et al. 1999), which is somewhat more extended and curved
towards the NW. 

\subsection{NGC 6814}

\begin{figure}[h]
\psfig{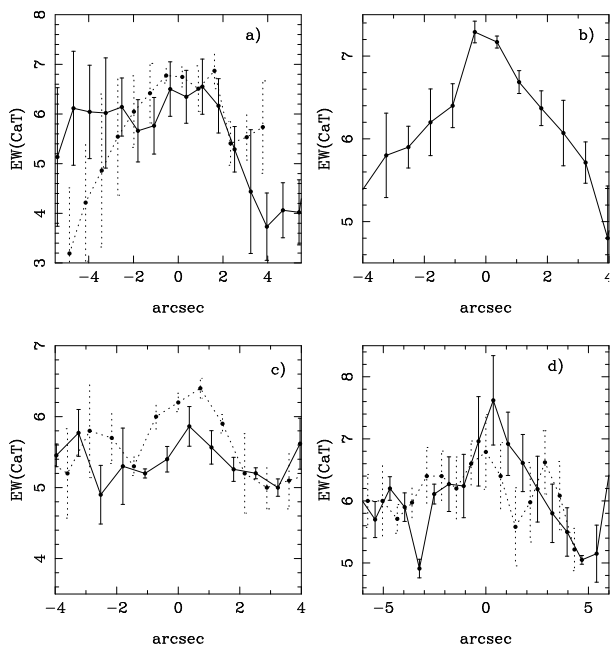}
\caption{Spatial distribution of the CaT equivalent width (EW) in the
central regions of: a) IC~184 (PA=7\degr~ as full line, PA=97\degr~ as dotted 
line), b) UGC~3223, PA=80\degr, 
c) NGC~2639 (PA=45\degr~ as full line, PA=135\degr~ as dotted line) and
d) NGC~6814 (PA=30\degr~ as full line, PA=120\degr~ as dotted line). 
The corresponding plot for NGC~6951 is given in P\'erez et al.  (2000).}
\label{ewCaT}
\end{figure}

NGC~6814 has a small inclination and it is therefore difficult to
derive its kinematical parameters accurately.
It is a single barred galaxy, with a $\pm$12 arcsec bar extending
along PA=25\degr~ (M\'arquez et al. 1999). Note 
that the innermost $\pm$3 arcsec show velocity dispersions about 30 \kms lower than the average values out to about 8 arcsec along PA=120\degr,
an effect which also seems to be detected along PA=30\degr.
Higher S/N data are needed in order to confirm the presence
of this drop. Although noiser, this profile resembles that of NGC~6951,
both showing a broader central drop than the other three galaxies
in our sample.
As shown in
Fig. \ref{ewCaT}d, a slight enhancement in EW(CaT) has been
detected along the extension of the dip along PA=30\degr.

The HST sharp-divided image of NGC~6814 shows a small (r $<$ 2 arcsec)
several-armed spiral structure (in dark grey), with a clear elongation to
the West along PA=120\degr~ in the innermost 1.5~arcsec and a
dust lane along PA$\approx$ 60\degr~ within 1 arcsec (Fig.~\ref{fig:n6814}).

\subsection{NGC~6951}

\begin{figure}[]
\begin{center}
\psfig{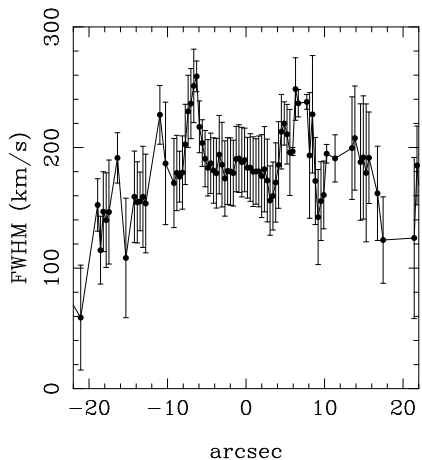}
\end{center}
\caption{NGC~6951. FWHM of the stellar component along PA=48\degr.}
\label{fig:n6951}
\end{figure}

The stellar velocity distributions of NGC~6951 are given in P\'erez et
al. (2000). We recall that two stellar components have been clearly
detected, with amplitude differences up to 50 \kms along PA=84\degr~ and
PA=138\degr. A similar behavior is observed along PA=48\degr~ in the region
between r=5 and 10 arcsec, which corresponds to stellar velocity
dispersion peaks (Fig.~\ref{fig:n6951}). Higher S/N data would be
required to obtain FWHM measurements for the two separate components,
but their physical connection with the FWHM peaks seems to be hinted
by our data. The enhancement of EW(CaT) in the central region was
already reported and discussed in P\'erez et al. (2000).

\section{Discussion and conclusions}

New stellar kinematics for a sample of four active galaxies
using the Calcium triplet have been obtained and we have 
shown that reliable stellar velocity curves can be traced.
Three of these galaxies show double bar-like confirmed structure
at optical and/or near IR frequencies. The previously reported data
for NGC 6951 (a single-barred galaxy with a nuclear spiral, P\'erez et
al. 2000) have been added to the present study, in particular the
velocity dispersion curve.

We have observed a dip in the stellar velocity dispersion in the
innermost $\pm$(1-3)~arcsec of these five active galaxies
based on observations made with 4~meter class telescopes. It is worth
noticing that the velocity dispersion drops have been detected in
three double-barred galaxies and in two single-barred ones, already
indicating that the presence of nested bars is not a necessary
condition; the drops in the later cases seem to be
broader than in the former.

 Note that such a dip is until now a rarely observed phenomenon,
reported so far only for eleven nearby galaxies (most of
them listed in Wozniak et al. 2003b; see references
below), three of them from recent VLT observations using the $^{12}$CO bands in
the near IR (Emsellem et al. 2001). This is most probably due to the
difficulty of estimating the FWHM of the stellar component with a
sufficient signal to noise ratio and spatial resolution.

Michel-Dansac \& Wozniak (2003) have recently modeled the evolution of
isolated galaxies over several Gyrs using a self-consistent N-body
code including stars, gas and star formation. With such models, they
succeeded in accounting for the drop in velocity dispersion observed
by Emsellem et al. (2001) in three double-barred and one
single-barred active galaxies, with a young stellar population born
from dynamically cold gas which is accreted in a circumnuclear disk
and creates the dip (Wozniak et al. 2003a, Wozniak \& Michel-Dansac
2003). We note that their simulations do not require the presence of
embedded bar structures, since such processes can be induced by the
primary bar itself (see also Wozniak et al. 2003b).

Since under these circumstances a luminosity dominating young stellar
component is expected in the region where the drops are observed, it
is worth trying to detect such a component in order to better validate
the model. We have attempted to do so by analyzing the equivalent
widths of the CaT lines along the dips and by inspecting the
inner structure in high resolution HST images.  The analysis of HST
images for the target galaxies shows that structure is indeed 
present in the regions where the dips occur in IC~184, UGC~3223 and
NGC~2639\footnote{With our small sample (5 galaxies) we find no
dependence of the drop sizes with respect to those of the outer/inner
bars; a larger sample should be needed in order to analyze this
point.}. The same conclusion is valid when inspecting the
sharp-divided frames of those galaxies with available HST archive
images
and with velocity dispersion drops reported in the literature, namely
NGC~1365 and NGC~1808 (Emsellem et al. 2001; for the Sy1 NGC~1097
there are only F218W images, that do not allow to detect such
structures), NGC~6503 (Bottema 1993), NGC~3627, NGC~4303, NGC~4579
(H\'eraudeau \& Simien 1998), NGC~4725 (H\'eraudeau et al. 1999;
diffraction spikes impede a reliable determination of the hinted
central elongated structure), NGC~4477 (Jarvis et al. 1988), NGC~3623
(de Zeeuw et al. 2002) and NGC~3412 (Aguerri et al. 2003). Note
that the images are at a wavelength range that does not include the
CaT, so we do not expect to find exactly the same structure. Also the
spectroscopic structure is diluted because of the poorer ground-based
spatial resolution. Therefore, the presence of structure in HST images
at the scales of the velocity dispersion drops is only a consistency
check.  The existence of a direct correlation between the two observables
needs further investigation with similar spatial resolution data.

The velocity dispersion drop in NGC~6814 seems to be as
broad as that in NGC~6951, and it is only for NGC~6814 and NGC~6951
(LINER/Sy2) that the drops seem to be present with no 
associated elongation in the images. This may indicate that
such structures either are seen face-on in these cases or are
absent. For the other galaxies, if the elongations are interpreted as
intrinsically circular disks, these disks would be placed almost
perpendicular to the main disk in NGC~2639 (i$_{inner~disk}$
$\approx$ 74\degr), at intermediate angle for IC~184 (i$_{inner~disk}$
$\approx$ 40\degr) and coplanar with the main disk in UGC~3223.

The Calcium triplet equivalent width, EW(CaT), 
is expected to increase for younger populations. Unfortunately, the
comparison with stellar population models to test the age of such
a population is rather risky, due to the already mentioned problem with
the determination of the continuum bands which impedes a quantitative
comparison with other authors (for instance, the models developed by
Garc\'{\i}a-Vargas et al. 1998 require EWs higher than those we have
measured to invoke an important contribution of super-giant
stars). The variations on the values of the EWs can still be
compared but,
due to the accuracy achievable by using EW(CaT) measurements on our
data, the analysis is not straightforward since we expect the relative
differences in the EW(CaT) to be comparable to our accuracy (from 1 to
1.5 \AA), and consequently such variations can, strictly speaking,
correspond to a constant value. Nevertheless, a hint of a local enhancement in
the drop region is suggested by the data, 
a behavior that could be due to the presence in these
regions of a population different from that of the bulge.

To confirm or discard Wozniak et al.'s models the analysis of higher S/N
spectra in a much broader wavelength range is needed to constrain the
age of the observed stellar populations detected in the central dips.
On this respect, we notice that some of the galaxies show evidence for
the presence of an intermediate (NGC~3627 and NGC~6503, Gonz\'alez
Delgado et al., in preparation) or young (NGC~4303, Colina et
al. 2002) population in the regions where the dispersion drops take
place and where elongations are seen in the HST sharp-divided images.

The case of NGC~6951 is specially interesting since an alternative
explanation for the presence of the apparent drop within r=5 arcsec in
this galaxy would be the existence of two stellar kinematic
components. This appears as a possibility to be checked in the other
four galaxies with higher spectral resolution and S/N data.  It would
therefore be of particular interest to re-observe these five galaxies
with an 8~meter class telescope,
to define with better precision the
characteristics of the dip, in particular its disk nature (for which
several position angles are required), and also to observe a
larger sample of galaxies with and without an AGN. Note that among the
16 galaxies with reported drops -- 5 of them presented for the first
time in this work -- 15 host AGNs\footnote{Their nuclear
classification has been taken from NED and is given in Table \ref{tabhst}.}, 
either Seyfert~1  (3),
Seyfert~2 (7), LINERs (2), or intermediate type
LINER/Sy2 (3); the only galaxy with no evident signs of nuclear
activity is NGC~3412 (Ho et al. 1995). The DEGAS sample is
particularly well adapted for this purpose, since it is well defined
and comprises only isolated galaxies, where the complicating effects
of galaxy interactions are not present.

\begin{acknowledgements}

We thank the referee, Eric Emsellem, for his excellent report, which
has largely helped us improve the clarity of the
paper. I.M. acknowledges financial support from the IAA and the
Spanish Ministerio de Ciencia y Tecnolog\'{\i}a through a Ram\'on y
Cajal fellowship.  This work is financed by DGICyT grants PB93-0139,
PB96-0921, PB98-0521, PB98-0684, ESP98-1351, AYA2001-2089,
AYA2001-3939-C03-01 and the Junta de Andaluc\'{\i}a grant TIC-144.  We
acknowledge financial support from the Picasso Programme d'Actions
Integr\'ees of the French and Spanish Ministries of Foreign
Affairs. F.D. acknowledges financial support from CNRS-INSU for
several observing trips.  E.P. acknowledges financial support from the
CONACYT (M\'axico) grant 36132-E. J. Maza gratefully acknowledges
support from the Chilean Centro de Astrof\'\i sica FONDAP 15010003.
We acknowledge Ignacio Marrero for his collaboration in some data
acquisition and calibration. This research has made use of the
NASA/IPAC extragalactic database (NED), which is operated by the Jet
Propulsion Laboratory under contract with the National Aeronautics and
Space Administration.

\end{acknowledgements}

\end{document}